\documentclass{aa}
\usepackage{graphicx}

\newcommand{\nqc}{NGC~4435}
\newcommand{\nqh}{NGC~4438}
\newcommand{\vcc}{VCC~1040}

\newcommand{\hi}{\mbox{H{\sc i}}}
\newcommand{\hii}{\mbox{H{\textsc{ii}}}}
\newcommand{\ha}{H$\alpha$}
\newcommand{\nii}{\mbox{[N{\textsc{ii}}]}}

\newcommand{\msol}{\rm{M_\odot}}
\newcommand{\kmps}{km s$^{-1}$}

\newcommand{\fas}[2]{$#1\fs#2$}
\newcommand{\as}[2]{$#1\farcs#2$}
\newcommand{\am}[2]{$#1\farcm#2$}
\newcommand{\ky}{K95}

\newcommand{\fantomm}{{\texttt {\textsc{FaNTOmM}}}}
\begin{document}

\title{A Virgo high-resolution \ha\ kinematical survey: I. \nqh
\thanks{Based on observations collected at the European Southern
Observatory, Chile, program 69.B-0496, and at the Observatoire du mont M\'egantic, Qu\'ebec, Canada}}

\author{L. Chemin\inst{1,2},  V. Cayatte\inst{3}, C. Balkowski\inst{2}, P. Amram\inst{4}, C. Carignan\inst{1},
A. Boselli\inst{5}, C. Adami\inst{5}, M. Marcelin\inst{4}, O. Garrido\inst{2,4}, O. Hernandez\inst{1,4}, J. Boulesteix\inst{4}}

\institute{D\'epartement de Physique, Universit\'e de Montr\'eal, C.P. 6128, Succ. centre-ville, Montr\'eal, Qc, Canada, H3C 3J7
\and Observatoire de Paris, section Meudon, GEPI, CNRS UMR 8111 \& Universit\'e Paris 7, 5 Pl. Janssen, 92195 Meudon, France
\and Observatoire de Paris, section Meudon, LUTH, CNRS-UMR 8102 \& Universit\'e Paris 7, 5 Pl. Janssen, 92195 Meudon, France
\and Laboratoire d'Astrophysique de Marseille, 2 Pl. Le Verrier, 13248 Marseille, France
\and Laboratoire d'Astrophysique de Marseille, Traverse du Siphon-Les trois Lucs, 13012 Marseille, France}
\offprints{Laurent Chemin, \email{chemin@astro.umontreal.ca}}

\date{Received   / Accepted }

\abstract{ New \ha\ emission-line observations of  the Virgo cluster galaxy \nqh\ are presented.
Fabry-Perot interferometry data at an effective angular resolution of $\sim 2\arcsec$ are used 
to map the kinematics of the ionized gas in the galaxy.
For the first time we obtain a velocity field covering a large area in \nqh, 
much larger than that deduced from previous \hi\ and CO maps.
The kinematics of the extended, low surface brightness \ha\ filaments to the West of the galactic disk is discussed. 
We report on the discovery of a northern \ha\ structure which is
clumpier than the other filaments. Evidence for multiple spectral components through the data-cube are presented  in
a nuclear shell  and in the approaching half of the disk.
The role of \vcc, a dwarf elliptical galaxy located to the South of \nqh, is presented to investigate
the origin of a small-scale stellar tail of \nqh. It could be due to a minor tidal interaction between the two galaxies.

 -- 
Instrumentation: interferometric -- Techniques: high angular resolution}

\authorrunning{L. Chemin et al.}
\titlerunning{A Virgo \ha\ kinematical survey: I. \nqh}

\maketitle

\section{Introduction}

The role of environmental effects is crucial for the evolution of galaxies.
Galaxies undergo  several different processes that directly affect their interstellar medium,
modifying their morphology and perturbing their dynamics, particularly while
crossing high density environments such as galaxy clusters.
 Well-known examples of such processes are tidal perturbations between galaxies 
 (Moore et al. 1996, Gnedin 2003) and ram pressure stripping exerted by the hot intracluster medium (hereafter
ICM, Gunn \& Gott 1972).

The Virgo cluster galaxy \nqh\ (VCC 1043) is a prototype of a disk interacting within
a dense environment. Its main properties are listed in Tab.~\ref{galax}.
It is one of the closest objects to the cluster core at only $\sim$ 1\degr\ in projection (or $\sim$ 280 kpc) from M87 and $\sim$ 28\arcmin\ (140 kpc) from M86.
Its morphology is the most perturbed among the Virgo galaxies, showing prominent stellar tails from North-East (NE) to South-West (SW)
of the disk  and a huge dust lane West of the disk (Fig.~\ref{imtot}),
molecular and neutral gas clouds displaced up to
$\sim 1\arcmin$ (4.7 kpc) West of the disk (Combes et al. 1988; Cayatte et al. 1990),
very extended (up to $\sim 2\arcmin$ or 9.4 kpc) ionized optical and X-ray emitting gas filaments
(Kenney et al. 1995, hereafter referred to \ky; Machacek et al. 2004) and an elongated region of $1.4$ GHz
continuum emission  (Hummel et al. 1983). It is  one of the most \hi\ deficient disks in the cluster (Cayatte et al. 1990, 1994). 
Numerical simulations succeeded in explaining the  NE and SW tidal tails formation 
 by a past high-speed tidal encounter with \nqc, a northern companion  (Combes et al. 1988).  
While a role for ram pressure stripping  by the ICM was proposed to explain some observations (Kotanyi et al. 1983, Cayatte et al. 1990, Keel \& Wehrle 1993),
 \ky\ speculated that the off-plane filamentary components could be regarded as consequences of an ISM-ISM interaction with \nqc.
 
In addition to this strongly disturbed large-scale morphology,
\ha+\nii\ WFPC2-\emph{HST} imagery (Kenney \& Yale 2002) and \emph{Chandra} X-ray observations  (Machacek et al. 2004)
have revealed gaseous nuclear bipolar shells. The origin of these nuclear shells could be due to 
an AGN (Machacek et al. 2004).

Optical Fabry-Perot (FP) interferometry is  quite appropriate to study the kinematics of \nqh\
since the ionized gas is the only gaseous component covering both  the inner disk regions as well as the outer ones.
As part of a survey dedicated to the study of the kinematics of Virgo Cluster galaxies
(Chemin et al. 2004, Chemin et al. in prep.), FP interferometry observations
of the \ha\ emission-line in \nqh\ are used to map its complete kinematics for the first time.
Section~\ref{fpres} presents the high resolution velocity field
of the galaxy, which reveals the kinematics in the disk, in the nuclear shells
and along the external filaments.
In Section~\ref{inter}  the kinematics of the filaments as well as the role of ram pressure stripping  in their evolution are discussed.
Section~\ref{inter} also proposes that a  minor interaction with the dwarf elliptical galaxy companion \vcc\ could be at the origin 
of morphological disturbances observed in the southern region of \nqh.

To be consistent with \ky, a distance to the Virgo cluster of 16 Mpc is adopted throughout the article.

\section{Observation and results}
\label{fpres}
\subsection{Data acquisition and reduction}
\begin{figure*}
\begin{center}
\includegraphics[width=18cm]{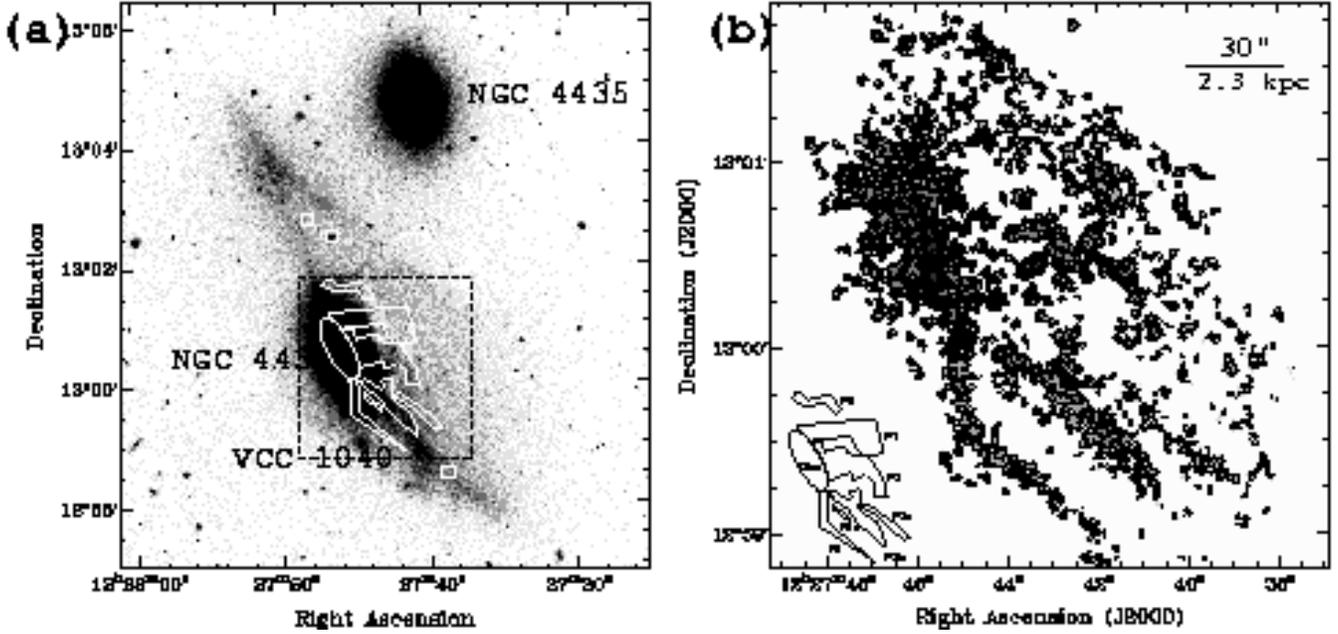}
\end{center}
\caption{\textbf{(a)}. A view of NGC 4435, NGC 4438 and VCC 1040. The optical map is a $B$-band image taken
from the GoldMine database (Gavazzi et al. 2003). Contours  display the approximate coverage of the total \ha\ emission.
White squares correspond to the positions of selected spectra from the M\'egantic FP observation (see \S\ref{fpres2}, 
Tab.~\ref{tabpvd} and Fig.~\ref{fspec}). The coordinates are in J2000. Notice the large dust layer on which the \ha\ emission
superposes to the North and West of the disk of NGC 4438. A dashed box corresponds to the field of view seen in the right panel.  
\textbf{(b)}.  Contours and grey-scale (displayed using a logarithmic stretch) map of the total integrated \ha\ emission.
The first contour is at a higher intensity  than the minimum level of the grey-scale image. 
The bottom-left, reduced-scale cartoon indicates the nomenclature used for the filaments.}
\label{imtot}
\end{figure*}
One set of observations of \nqh\ was performed in April 2002 at the 3.6m European Southern Observatory telescope equipped
with a focal reducer, a scanning FP interferometer and the photon counting camera \fantomm
\footnote{\fantomm\ stands for Fabry-Perot de Nouvelle Technologie de 
l'Observatoire du mont M\'egantic. See http://www.astro.umontreal.ca/fantomm for more details.} (Gach et al. 2002).
Table~\ref{tabobs} describes the characteristics of the observations. 

A \fantomm\ \ha\ emission-line observation obtained  at the 1.6m Observatoire
du mont M\'egantic (OmM) T\'elescope was also used to map a large field of view around \nqc\ and \nqh\ (see Tab.~\ref{tabobs}).
The observation is used to search for emission-lines outside the disk in the
large-scale tails of \nqh. Although the OmM data-cube has low signal-to-noise ratio,   the detection of ionized gas was successful
in the disk and the brightest parts of the filaments, as well as at three locations  represented by white squares in Fig.~\ref{imtot} (see also Tab.~\ref{tabpvd}).
Only the spectra at these three locations  are presented for this observation (see section~\ref{fpres2}). 

\begin{table}
\caption[ ]{Basic properties of \nqh\ (VCC 1043).}
\begin{tabular}{lc}
\hline
 Right ascension (J2000)& 12$^h$ 27$^m$ 45.6$^s$ \\
 Declination (J2000)&  +13$\degr$ 00\arcmin\ 32\arcsec \\
 Type$^1$ & Sb (tides)\\
 Nucleus$^2$ & LINER 1.9\\
 Distance  &  16 Mpc\\
 Linear scale &  78 pc arcsec$^{-1}$\\
 Heliocentric systemic velocity$^3$ $v_{\rm sys}$ &  71 \kmps\\
 Inclination$^4$ $i$ &  87\degr \\
 Disk position angle$^3$ $P.A.$ &  29\degr \\
\hline
\end{tabular}
\begin{itemize}
\item[$^{\rm (1)}$]   Morphological type from Binggeli et al. (1985)
\item[$^{\rm (2)}$]    Ho et al. (1997)
\item[$^{\rm (3)}$]    $v_{\rm sys}$ and $P.A.$ from Kenney et al. (1995)
\item[$^{\rm (4)}$]    $i$ from LEDA
\end{itemize}
\label{galax}
\end{table}

\begin{table}
\caption[ ]{Observational setups.}
{\tiny
\begin{tabular}{lll}
\hline
 Telescope & ESO 3.6m & OmM 1.6m\\
 Observation date &   2002, Apr., 6$^{\rm th}$ &  2004, Feb., 24$^{\rm th}$ \\
 Field of view &  \am{3}{6} &  \am{13}{7} \\
 Data-cube size &  \tiny $512\times512\times24$ &  $512\times512\times48$\\
 ($X$,$Y$) pixel size &  \as{0}{42} &  \as{1}{61}\\
 Interference order @ \ha &  793 &  765\\
 Free spectral range  @ \ha\ (\kmps) &  378 &  392  \\
 Channel width (\kmps) &  15.75 &  8.16 \\
 Scanning wavelength (\AA) &  6564.3 &  6561.7 \\
 Interference filter central $\lambda$ (\AA) &  6567.7 &  6565.0 \\
 Interference filter FWHM (\AA) &  12 &  30  \\
 Exposure time (s) &  9360  &  14400 \\
 Average seeing & \as{0}{8}  & \as{2}{5} \\
\hline
\end{tabular}}
\label{tabobs}
\end{table}

\subsection{\ha\ Fabry-Perot emission-line and velocity maps}
\label{fpres2}

The basic pre-processing data reduction steps of raw FP data-cubes in order to obtain a calibrated data-cube 
have been described elsewhere (see e.g. Chemin et al. 2003).
Once calibrated, the  integrated emission of the galaxy corresponds to the sum of the signal of all channels above a continuum threshold.
The threshold is chosen so that 15\% of all the channels are under the  continuum level, and the remaining 85\% channels are consequently due to
the \ha\ emission-line. We call this emission the total integrated emission. The pixel velocity corresponds to the barycentre of the emission-line. 
The case of a peculiar region for which some of its pixels exhibit two lines is described below. For these pixels, the velocity is not unique and
 a specific treatment must be done (see ``Southern disk-Filament F4 region" paragraph).

\begin{figure*}[!th]
  \centering
 \sidecaption
 \includegraphics[width=11.5cm]{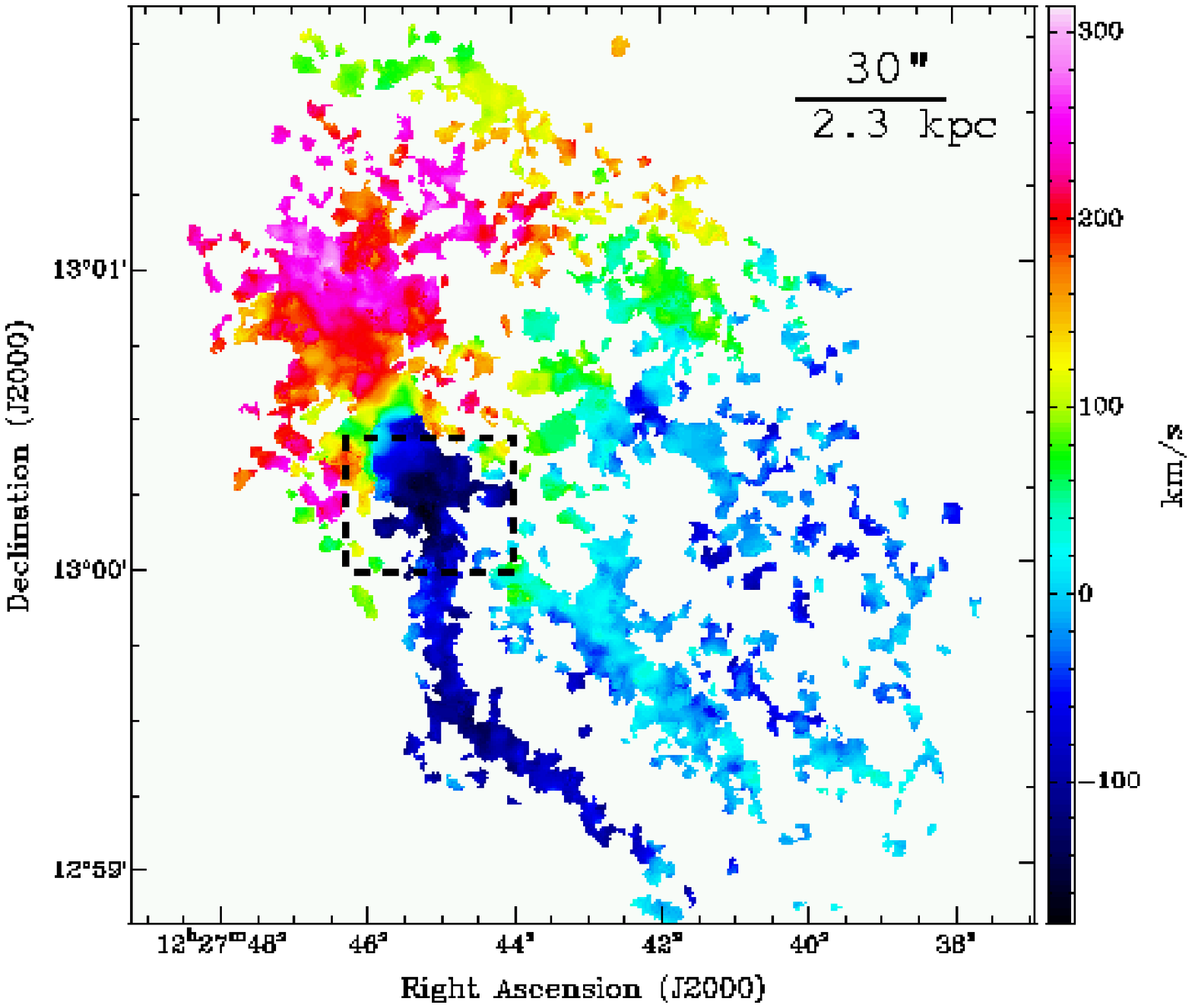}
 \caption{~\ha\ velocity field of NGC 4438. Radial velocities are respect to the heliocentric rest-frame. 
 A dashed box corresponds to a region where two velocity components are detected in \emph{some} of its pixels (see section~\ref{fpres2}, 
 paragraph ``Southern disk-Filament F4 region").  The  most blue-shifted velocity values are shown for these pixels. 
 The location of these pixels is shown in Fig.~\ref{vf2}a and their most red-shifted velocity values in Fig.~\ref{vf2}b.}
\label{vf1}
\end{figure*}

Figure~\ref{imtot}b and Figure~\ref{vf1} present the total integrated \ha\ emission-line and the velocity field 
of \nqh\ obtained using a data-cube appropriately smoothed to effective resolutions between
\as{1}{26} and \as{2}{52} (corresponding to 3 and 6 pixels FWHM) in order to increase the signal-to-noise ratio in the field-of-view. 
Our \ha\ map is in good agreement with known deep \ha\ imagery (\ky; Kenney \& Yale 2002; Gavazzi et al. 2003). 
The \ha\ velocity field 
is in good agreement with the few  long-slit measurements (\ky) where both studies coincide.
The most striking feature  is the filamentary morphology extending
 West and South  up to \am{2}{1} (or $9.8$ kpc) from the nucleus
and on which are superposed a few brighter clumps, probably \hii\ regions.
Some of these structures have already been described (\ky, but see also Keel \& Werhle 1993)   
and similar names as in \ky\ for the filaments will be used hereafter, as shown in the bottom-left corner of Fig.~\ref{imtot}b.
In the following, the kinematics of the different components of the galaxy are described: disk, nuclear shells, filaments and 
new found structures.

 \begin{figure}[!]
 \centering
 \includegraphics[width=9cm]{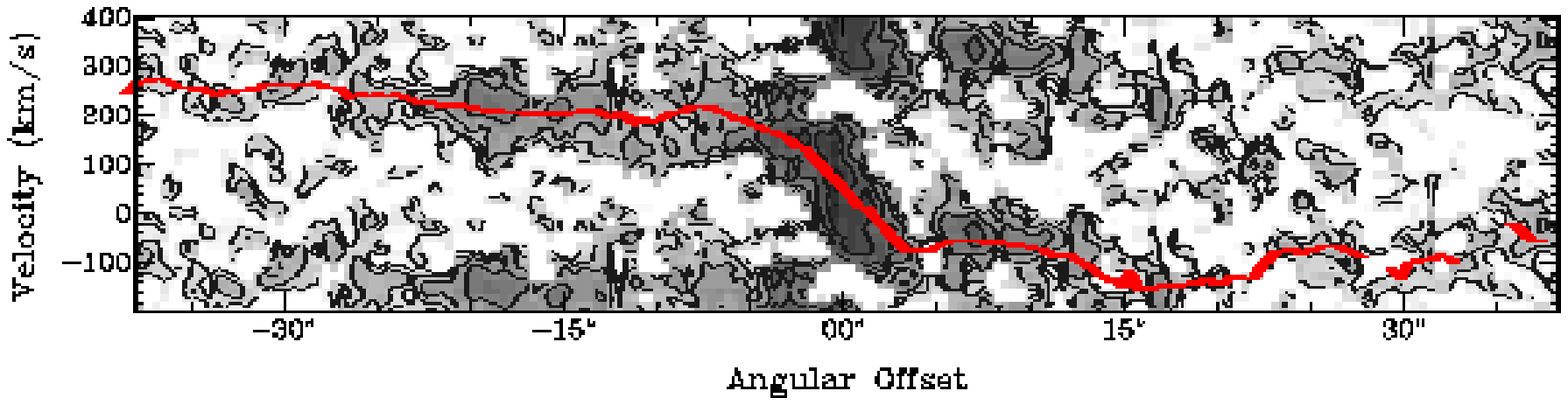}
 \includegraphics[width=6cm]{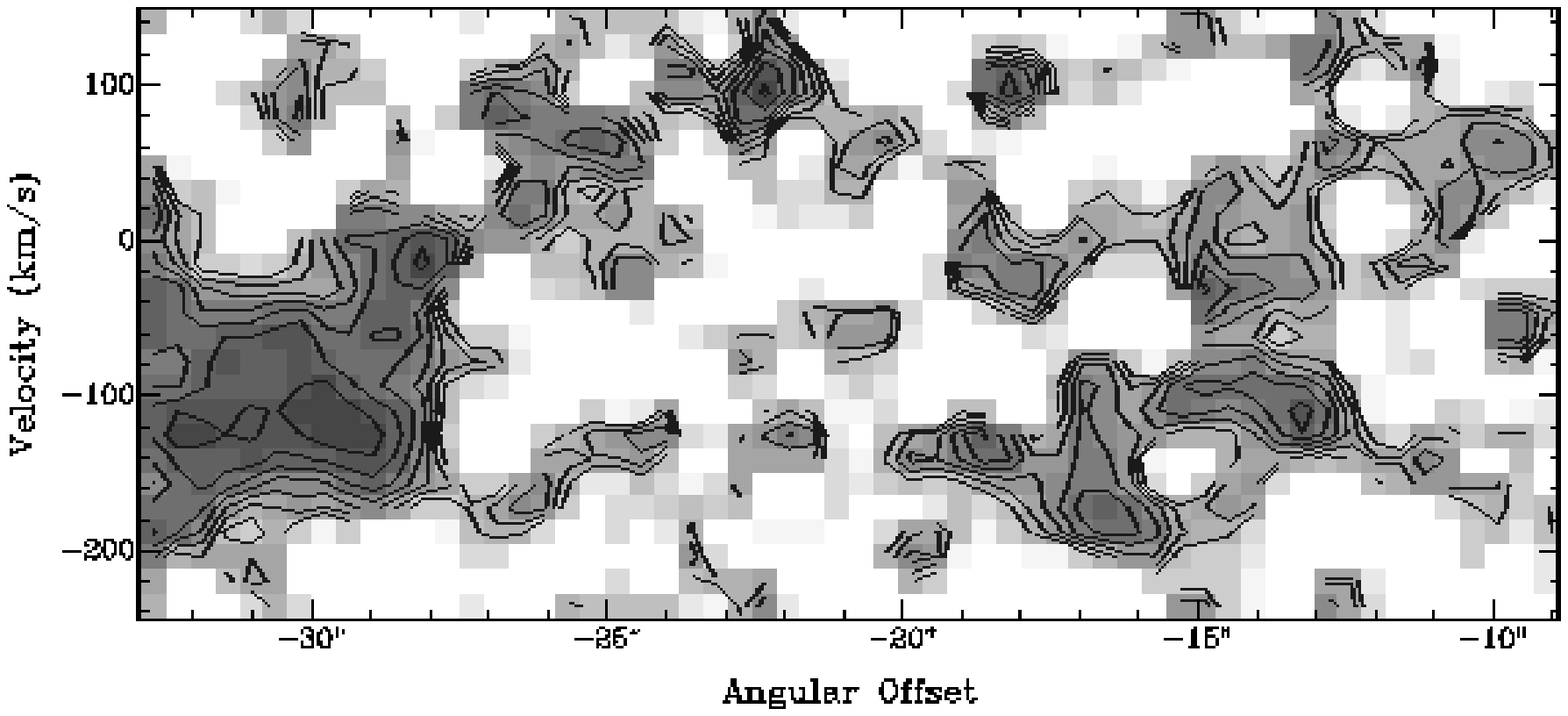}
\caption{~\ha\ Position-velocity diagrams through different regions of the continuum-free data-cube. Top : along the major axis of the disk ($P.A. = 29\degr$).   
Bottom  : along the Filament F4 ($P.A. = 180 \degr$). 
 A logarithmic grey-scale and contoured stretch is used for the \ha\ emission-line.  A red solid line  corresponds to 
 the velocity field values (Fig.~\ref{vf1}).  The centre and orientation of the slices are listed in Tab.~\ref{tabpvd}. 
 The effective angular resolution of the data-cube used for the diagrams is \as{0}{84} (or 2 pixels FWHM).}
 \label{pvd1}
\end{figure}

\begin{figure}[!t]
  \centering
\includegraphics[width=\columnwidth]{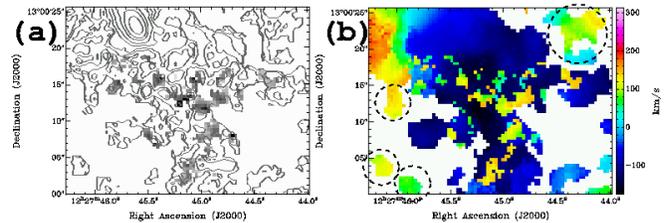}
 \caption{Integrated emission (a) and velocities (b) of the most redshifted line for peculiar pixels in the dashed box region of Fig.~\ref{vf1} 
 (see section~\ref{fpres2}, paragraph ``Southern disk-Filament F4 region" for details). 
 \textbf{(a)}. The grey-scale pixels display  the emission integrated under the most redshifted emission line  
 of these peculiar pixels  and contours display the total integrated emission of the galaxy (already seen in Fig.~\ref{imtot}b).
   \textbf{(b)}. Same velocity field as in Fig.~\ref{vf1} but this time showing the red-shifted velocity component for the peculiar pixels. 
   Notice that these redshifted lines have velocities in contrast with that of the approaching half of the disk but 
    in agreement with those of structures outside the disk and Filament F4 (encircled by dashed lines).}
\label{vf2}
\end{figure}

\emph{\textbf{Disk}}.--
In the disk, the gas emission is dominated by   few \hii\ regions South and North from
the nucleus   and by a nuclear shell (see next paragraph).
No attempt to derive an optical rotation curve was undertaken due to a highly inclined disk,
obscured by dust lanes, that would cause large uncertainties in the  circular velocities.  
The average minimum velocities are down to $-150$ \kmps\ in the southern approaching
half of the disk.
Our results are consistent with the CO observations which show an inner molecular disk with velocities  lower than $- 100$ \kmps\
(see Figs. 10 and 12 of \ky, see also discussion in \ky\ about the Combes et al. observations).
  
 Figure~\ref{pvd1}  represents the position-velocity diagram along the major axis of the galaxy ($P.A. =$ 29\degr) and shows
a classical rotating disk. The \emph{kpvslice} task of the \textsc{karma} package (Gooch 1995) was used to make the position-velocity diagram.
The pseudo-slit width is between 3 and 4 pixels (i.e. between \as{1}{3} and \as{1}{7})
with a pixel weighting  inversely proportional to the distance of the slice.
For clarity reasons, a red solid line,  which represents   the  velocity field values (Fig.~\ref{vf1}), is drawn to follow the
main interference order since more than one order is shown here. Lines are broader in the centre than in any other regions of the major axis.
 
\emph{\textbf{Nuclear shells}}.-- 
Kenney \& Yale (2002) found evidence for a bipolar nuclear shell in \nqh\ from \ha+\nii\ WFPC2-\emph{HST} images. 
The NW and SE shells are intrinsically  asymmetric in luminosity and length, in both optical and X-ray emission-lines and
radio continuum (Kenney \& Yale 2002, Machacek et al. 2004). Using an emission-line velocity dispersion
measurement along the western edge of the NW shell, Kenney \& Yale (2002) claimed that the shells are caused by a nuclear outflow in \nqh.
Figure~\ref{imtot}b shows that the NW shell is one of the brightest structure that is observed in the disk (R.A.$\sim$ 12$^h$ 27$^m$ 45.7$^s$, Dec.$\sim$ +13$\degr$ 00\arcmin\ 33\arcsec)
 while the SE shell is indeed fainter (R.A.$\sim$ 12$^h$ 27$^m$ 46.5$^s$, Dec.$\sim$ +13$\degr$ 00\arcmin\ 26\arcsec). 
 Figure~\ref{vf1}  shows that most of the gas motions in the SE shell is at  $v_{\rm obs} \sim 75 \pm 5$ \kmps,  which is close to the systemic velocity ($71$ \kmps).
At the tip of the shell, the velocities reach $\sim 115$ \kmps.

 \begin{table}
\caption[ ]{(Top) Position-Velocity slice parameters along the major axis and Filament 4 from the ESO FP-observation. 
(Bottom) Position of selected spectra in the OmM FP-observation.}
\begin{tabular}{lll}
\hline
 \textbf{ESO~:} &~ &~\\
 \textbf{Slice} & \textbf{Centre coordinates} & \textbf{$P.A.$} \\
                & \textbf{(J2000)} & \textbf{(\degr)} \\
 Major axis &  12$^h$ 27$^m$ \fas{45}{698} +13$\degr$ 00\arcmin\ \as{31}{38} & 29\\
 Filament 4 &  12$^h$ 27$^m$ \fas{44}{997} +12$\degr$ 59\arcmin\ \as{49}{57} & 180\\
\end{tabular}
\begin{tabular}{lll}
\hline\hline
 \textbf{OmM~:} &~ &~\\
 \textbf{Spectra} & \textbf{Centre coordinates} & \textbf{Velocity}\\
                  & \textbf{(J2000)} & \textbf{(\kmps)}\\
 South (a) &  12$^h$ 27$^m$ \fas{38}{887} +12$\degr$ 58\arcmin\ \as{32}{73} & -35  \\
 North (b) &  12$^h$ 27$^m$ \fas{46}{410} +13$\degr$ 02\arcmin\ \as{29}{27} & -60  \\
 North (c) &  12$^h$ 27$^m$ \fas{48}{000} +13$\degr$ 02\arcmin\ \as{45}{00} & -200 \\
 \hline
\end{tabular}
\label{tabpvd}
\end{table}

\begin{figure*}
 \sidecaption
\includegraphics[width=4.5cm]{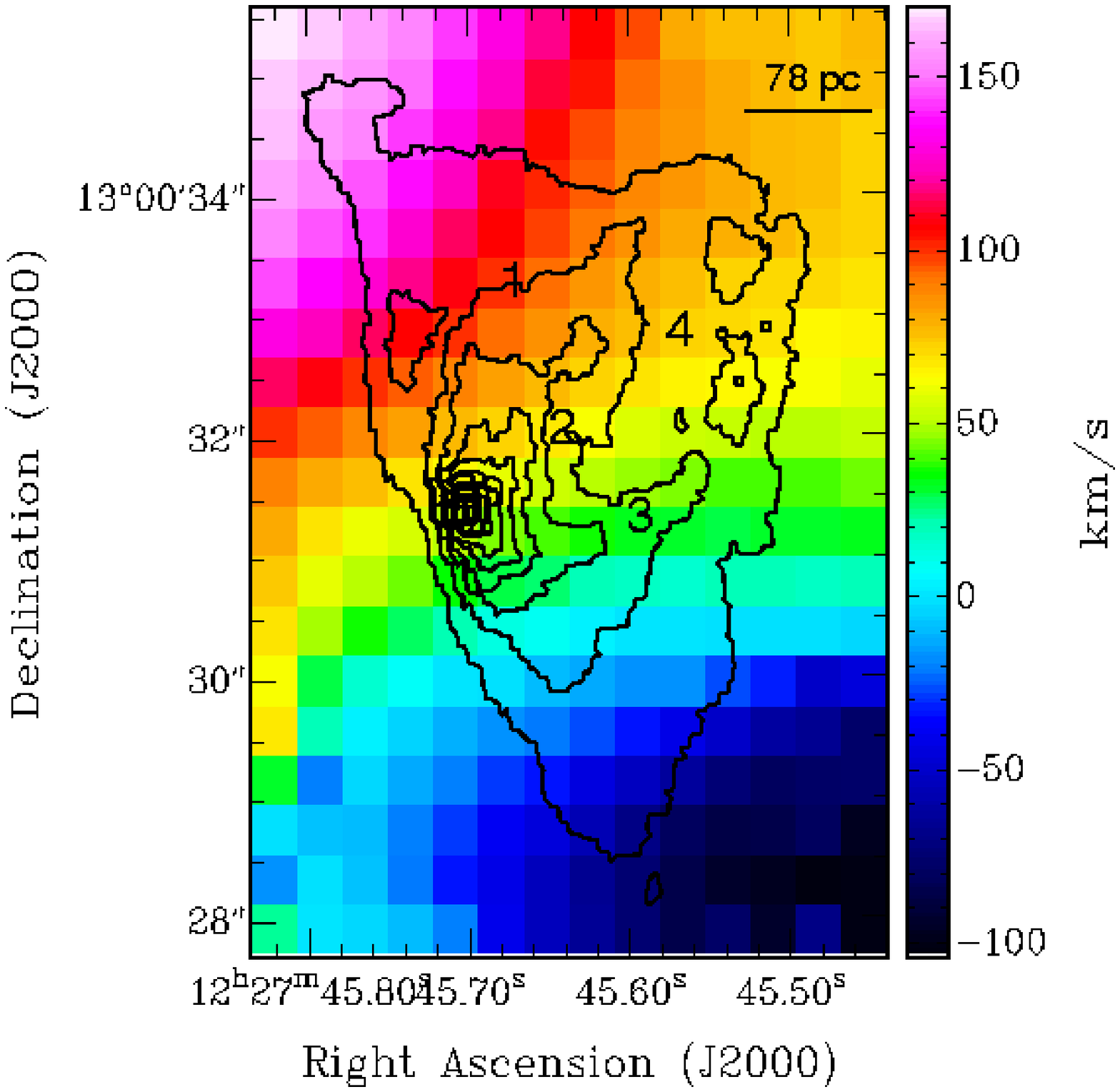}\includegraphics[height=4.0cm]{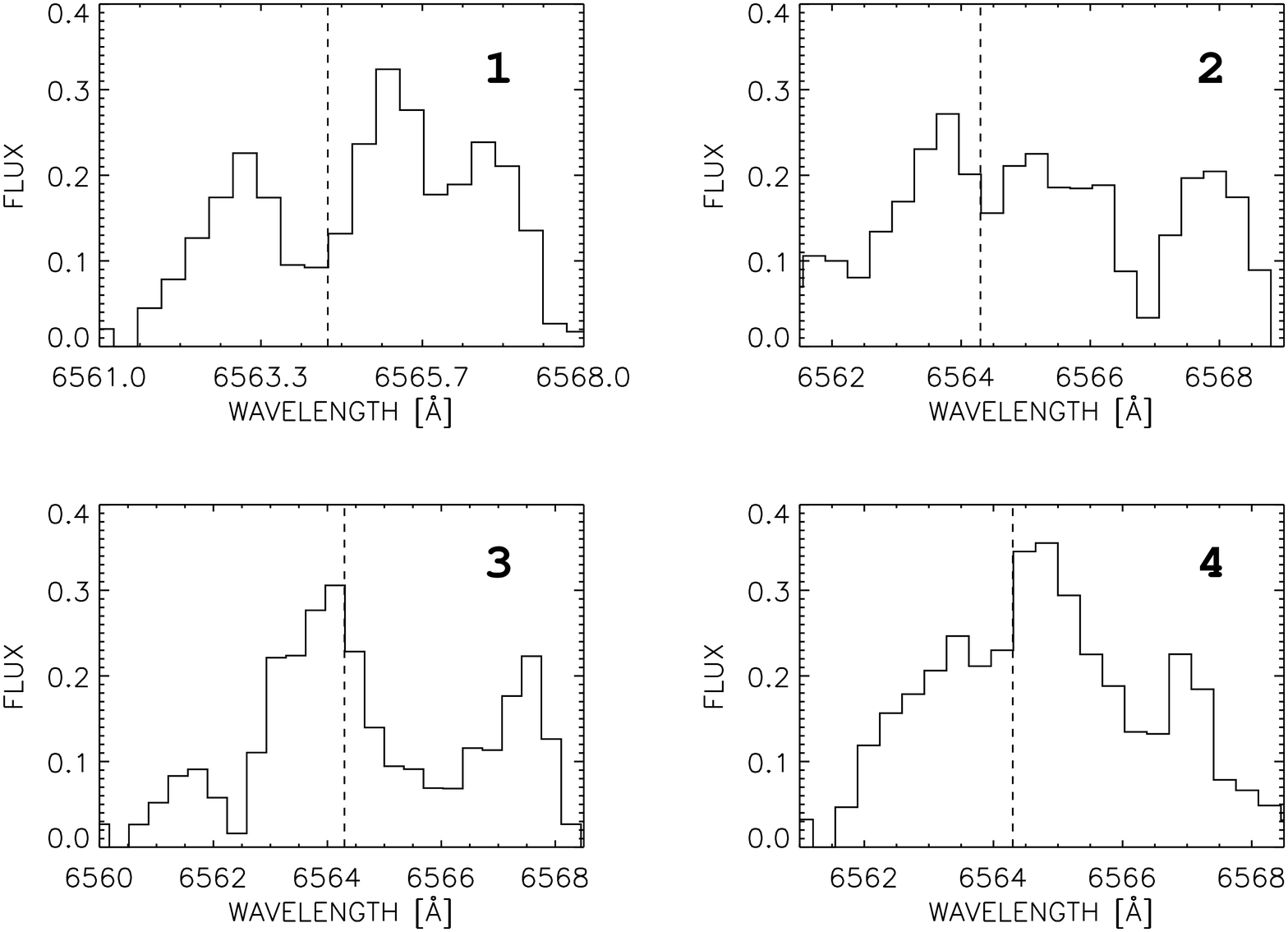}
\caption{~\ha~+~\nii~ WFPC2-\textit{HST} contours of the NW nuclear shell (Kenney \& Yale 2002) superposed on the  corresponding
 \ha\ FP coloured velocity field. The spatial resolution of the \ha~+~\nii~ HST image is \as{0}{1}.
 The 78 pc segment corresponds to 1\arcsec\ (D=16 Mpc). Numbers 1 to 4 refer to the location of
 selected ESO-FP normalized continuum-free spectra along the shell (right panels).
 The flux unit of the spectra is arbitrary. The dashed lines indicate the systemic velocity of the galaxy. }
 \label{fnuc}
\end{figure*}

 In Figure~\ref{fnuc}, we focus on the NW shell kinematics. The shell has radial velocities increasing from $\sim -60$ \kmps\
to $\sim 140$ \kmps\ (from SW to NE of the shell, respectively). We use here the high resolution  \ha~+~\nii~ WFPC2-\textit{HST}
  images (Kenney \& Yale 2002) taken from the \textit{HST}  archive and reduced as explained in \S\ref{inter2}.
The velocities are close to the systemic velocity, principally along  the apparent  semi-major axis of the shell.
As seen in spectra extracted from the \emph{full angular resolution} data-cube and selected at different positions along the shell (right-panel diagrams of Fig.~\ref{fnuc}),
the profiles are relatively wide in the shell.
The restricted free spectral range of 380 \kmps\ of the ESO-FP observation does not allow us to
 derive the velocity dispersions because the bases of the profiles are missing. 
The spectra show several velocity components.  The major component in each spectrum
 is at a velocity given by the velocity field, close to the systemic velocity of the galaxy of $\sim 70$ \kmps.
 A redshifted (blue-shifted, respectively) component is detected at  velocities  up to $\sim 150$ \kmps\ (down to $\sim 50$ \kmps) relative to the systemic velocity.
 The blue-shifted component is detected in diagrams 1 to 3, but hardly in diagram 4.  The reader should be aware that these
 additional components are known $\pm 380$ \kmps, i.e. the ESO interfringe, because there is no direct 
 way to determine the interference order at which they are emitting.  
  Determining their true velocity would need larger spectral band observations than ours.
 Line splittings in nuclear shells often indicate an expanding bubble (Veilleux et al. 1994).
 This is here the first direct evidence that the shell is expanding.
  An accurate determination of the intrinsic outflow velocity and velocity dispersion would
require a coupled kinematical and geometrical model of the shell  (see e.g. Veilleux et al. 1994, for NGC 3079),
which is beyond the scope of the article.

\emph{\textbf{Filament F0}}.-- This northern filament was not previously described in \ky\
although present in their data.
 It is located at $65$\arcsec\ (5.1 kpc projected)  North
 from the nucleus and is mostly traced by a string of \ha\ knots (probably \hii\
regions) embedded in faint diffuse gas, running nearly parallel to the EW direction.
Its morphology thus differs from that of the other filaments, which are mostly only diffuse.
It has a length  of 45\arcsec\ (3.5 kpc) although it could be longer toward the SW  and connected to Filament F1.  
Compared with a  $B$-band  image of the galaxy (Fig.~\ref{imtot}), F0 exactly superposes on the northernmost part of the large
western dust layer.
Its velocities ($\sim 80$ \kmps) are of the same order as the systemic velocity of \nqh\ but admit a maximum of
$\sim 110$ \kmps\ at its western tip - the \hii\ clump at R.A.=12$^h$ 27$^m$ 44.4$^s$, Dec.=+13$\degr$ 01\arcmin\ 36\arcsec. 
The velocity of this latter clump is consistent with that  of a \hi\ counterpart at 120 \kmps\ (Cayatte et al. 1990) as well as
the one obtained with the long-slit observation  (\ky, their $P.A.= 0\degr$ slit).
 A comparison of our \ha\ map with the  \emph{Chandra} X-ray image (Machacek et al. 2004) shows that  F0 has no X-ray emission  counterpart. 
 Its evolution appears slightly different from that of the other filaments.

\emph{\textbf{Filament F1}}.-- A velocity gradient is observed along part of the filament over a distance of 20\arcsec\ (1.6 kpc) from 250 \kmps\
down to about the systemic velocity of 71 \kmps\ (from East to West respectively).
One may  notice that the western tip of F1 actually presents two regions with distinct velocities: a northern part at
$v_{\rm obs} \sim 130$ \kmps\ (R.A.$\sim$ 12$^h$ 27$^m$ 41.6$^s$, Dec.$\sim$ +13$\degr$ 01\arcmin\ 10\arcsec)
and a southern one at the systemic velocity.
In fact, this northern part has a velocity comparable with that of the western extremity of F0 (within 20 \kmps)
and  could be a genuine extension of F0.
Combes et al. (1988) reported wide CO-line profiles at a velocity of $\sim 85$ \kmps\ in the molecular
cloud located at 1\arcmin\ (4.7 kpc) NW from the nucleus. Taking into account
the resolution of 23\arcsec\ for the CO observations,  it is probable that these two regions at significantly
different velocities  could not be resolved by the \emph{IRAM} beam and thus artificially enlarged the CO emission-lines.
These latter are then nearly centered at the velocity of the highest surface brightness component.

\emph{\textbf{Filament F2}}.--   This filament also presents a  velocity gradient, with velocities 
 between $\sim -15$ \kmps\ at its western tip and  $\sim 200$ \kmps\ at the apparent connection with the disk.
Compared with the largest neutral hydrogen concentration detected in the \emph{VLA} observations (Cayatte et al. 1990),
the  \ha\ and \hi\ velocities are very similar  for the regions of overlap (60-80 \kmps).

\emph{\textbf{Filament F3}}.-- It is composed of three distinct, nearly parallel substructures F3a, F3b and F3c.
F3b represents the main body of F3 and exhibits an average velocity of $\overline{v}_{\rm obs} \sim 20$ \kmps\ occurring in its brightest parts, although
lower velocities down to $\sim -30$ \kmps\ are also observed along it,
as well as higher ones (up to $\sim 70$ \kmps) at its eastern tip, close to F4. F3b actually presents  wide
 profiles (up to 165 \kmps\ FWHM, corrected for the instrumental Airy profile) which could be due to turbulent motions.

F3a has negative velocities ($\sim -25$ \kmps)
to its NE side and null velocities (on average) on the other side.
A new, short filament (F3c) is detected ($\sim 15$ \arcsec, 1.2 kpc), barely seen and not discussed in \ky.
It appears to be linked to F3b by a $\sim$ 5\arcsec\ (400 pc) ionized gas structure.
F3c appears to be an extension of a bright warped stellar structure (see Fig.~\ref{finter}a)
and mostly shows velocities of $-20$ \kmps, but with a drop down to $-80$ \kmps.

\emph{\textbf{Filament F4}}.--
F4 appears to be the longest filament with a projected length of at least \am{1}{5} (7 kpc).
Its first part originates in the southernmost side of the  disk and
is along the NS direction, running parallel but clearly offset from a stellar tidal tail (see \S\ref{inter2} and Fig.~\ref{finter}a).
Its second part is directed along the NE-SW direction and is almost parallel to F3.  

F4 is the most blue-shifted filament w.r.t. the systemic velocity, presenting
an average and nearly constant velocity of $\overline{v}_{\rm obs} \sim -85$ \kmps.
F4 may contribute to part of the kinematics in the southernmost parts of the disk but it is obviously less blue-shifted
than further inside in  the disk, where velocities down to $-150$ \kmps\ are observed.
It also exhibits a velocity drop (down to $\sim -135$ \kmps) occurring in its bending region.
Ionised gas at $\overline{v}_{\rm obs} \sim -85$ \kmps\ is also observed close to the bending region but outside F4.
The clump at the tip of F4 is at $\sim -5$ \kmps\ and therefore appears kinematically independent from
the whole filament.

\emph{\textbf{Southern disk-Filament F4 region.--}}
The dashed box of Fig.~\ref{vf1} corresponds to a particular region, where F4 appears to meet the SW half of the disk. 
Some of its pixels exhibit two emission lines (Fig.~\ref{pvd1}). 
For those pixels, a barycentre computed over the whole profile, as done for all other pixels of the field-of-view that have only one spectral line, 
would give incorrect velocities due to the presence of the two competing lines. Two velocities need to be computed for them.
 A blue-shifted component  has a velocity comparable with 
that of neighbouring pixels that show only one spectral component ($< - 100$ \kmps). It 
allows to follow a kinematic continuity in the SW approaching half of the disk. 
 Its pixels are thus shown on the velocity map of Fig.~\ref{vf1}.  
Figure~\ref{vf2} shows the pixels of the second, redshifted component (up to $\sim 100$ \kmps). 
A discontinuity is clearly seen with respect to the surrounding pixels showing only one emission line.
Its kinematics is hard to reconcile with the rotating, approaching, blue-shifted pattern of the SW half of the disk and of Filament F4. 
It does not suggest that the emission comes from the disk or Filament F4,  unlike the blue-shifted line. 
Its velocity is in better agreement  with that of diffuse \ha\ structures outside the disk and F4 (see circles in Fig.~\ref{vf2}).  
The pixels seem to connect these latter different redshifted structures, as if all the redshifted gas would belong to a same morphological structure. 
This latter could  be an extension of Filament F2 to the East.
The amplitude of the most redshifted line tends to be brighter than the blue-shifted one. 
However the transmission curve of the narrow interference filter may bias this claim (the filter transmission being 
higher in this part of the   spectral range than in its blue part) 
so that it is hard to discriminate which component really dominates the total integrated emission.   
The  integrated intensity map of the redshifted line is represented in grey-scale in Fig.~\ref{vf2}, with contours showing the 
total \ha\  intensity. This local map should only be used to locate the peculiar pixels. Further observations would be needed to  
confirm our new finding and to accurately map the intensity of the most redshifted component.

\emph{\textbf{Evidence for new structures}}.--
   An isolated clump is detected West of F0 (R.A.=12$^h$ 27$^m$ 42.5$^s$, Dec.=+13$\degr$ 01\arcmin\ 44\arcsec) with a radial velocity
of $145 \pm 5$ \kmps. It is barely seen in the \ha\ image of  Kenney \& Yale (2002) and no other molecular gas or \hi\ counterparts
have been observed at this location. The \ha\ image of Kenney \& Yale (2002) also shows two other isolated knots located at 60\arcsec\ and 65\arcsec\
NE from the nucleus and close to the eastern edge of their field of view. These knots do not present \ha\ emission in the FP data-cube but
typical continuum profiles and thus they could be foreground stars or more distant ojects.

To the West of the field of view, in a region centered at R.A. $\sim$ 12$^h$ 27$^m$ 39$^s$, Dec. $\sim$ +13$\degr$ 00\arcmin\ 00\arcsec, a very faint and patchy
emission is observed at velocities between $-80$ and $20$ \kmps. It is possible that
sky emission-line residuals contaminate this extremely  low surface brightness region.
Nevertheless, most of this emission can be seen in the Kenney \& Yale (2002) \ha\ image,
although not discussed, as well as in the \ha\ image of GoldMine (Gavazzi et al. 2003). Therefore the filamentary structure appears
much more complex and extended towards the West than the previously discussed linear filaments.

To the East of F4,  a new faint \ha\ structure is observed, with clumps at velocities between  70-90 \kmps\
(see e.g. the clump at R.A. $\sim$ 12$^h$ 27$^m$ \fas{46}{053}, Dec. $\sim$ +12$\degr$ 59\arcmin\ \as{54}{06}).
These \ha\ clumps superpose on a short  X-ray filamentary-like structure in the \emph{Chandra} observations of
Machacek et al. (2004), although not discussed by them. The \ha\ clumps
are separated from F4 by a  dust lane (see Fig.~\ref{finter}).
We think that we do not detect a filament like the X-ray one
because our FP profiles are at too low signal-to-noise in this region. For the present time,
the origin of the \ha\ clumps and of the short X-ray structure is unknown : they could have a relation with F4,
and/or with the nuclear outflow,  or perhaps with the pixels that show 
a redshifted spectral line in the southern disk region (see previous paragraph).

\begin{figure}[b!]
\includegraphics[width=9cm]{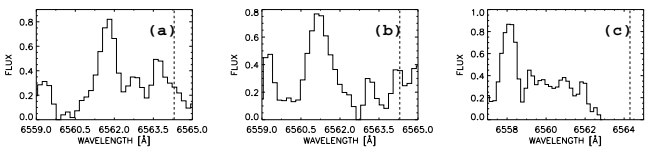}
\caption{~Fabry-Perot continuum-free spectra of selected regions (see coordinates in Tab.~\ref{tabpvd} and white squares in Fig.~\ref{imtot}).
The spectra are averaged in a box of $8\arcsec\times8\arcsec$ around centres of Tab.~\ref{tabpvd}. The flux unit is
arbitrary. The dashed lines indicate the systemic velocity of the galaxy.}
\label{fspec}
\end{figure}

Three clumps outside the ESO-FP field-of-view have been detected with the OmM observations. Their spectra is shown in
 Fig.~\ref{fspec} and their velocities listed in Tab.~\ref{tabpvd}. Their positions are marked by white squares in Fig.~\ref{imtot}.
The lines observed here cannot be associated to night sky emission-line residuals because these latter always present very sharp wings and
are relatively narrower than ionized gas emission-lines, which is not observed.
The left panel (Fig.~\ref{fspec}a) shows that ionized gas is  detected in the SW stellar tail, in the continuity of F3b and
 with a line at a velocity of $v_{\rm obs} \sim -35$ \kmps. This is comparable with the velocities observed along F3b.
On broad-band images, the knot associated with the middle panel spectrum (Fig.~\ref{fspec}b) appears as a system having two objects.
The right-hand spectrum (Fig.~\ref{fspec}c) corresponds to a blue clump on the RGB image of the GoldMine database.
These two latter clumps could be \hii\ regions associated with the northern stellar tail.
Their velocities appear  extremely blue-shifted, in contrast with the kinematics observed
 in the northern, receding half of the disk. However, because the velocity
 of any \emph{isolated} clump is known $\pm 390$ \kmps\ (i.e. the OmM observation free spectral range),
 it is reasonable to deduce true velocities of $+330$ \kmps\ and $+190$ \kmps\ for them,
 when adding an interfringe to the measured velocities.
 In this case, their velocities would be in better agreement with the kinematics in the North.
 Their velocities need to be confirmed by other spectroscopic measurements.

\section{Discussion}
\label{inter}

\subsection{Global kinematics of the filaments}
\label{inter1}

 Evidence for a high-speed tidal interaction ($\sim 100$ Myrs ago) with the northern
companion \nqc\ at the origin of the prominent NE and SW stellar tails and of part of 
the molecular gas above the galactic disk were first presented
in Combes et al. (1988). Furthermore, \ky\ proposed that the off-plane gaseous distribution could be due to 
an ISM-ISM collision  with \nqc\ and that  the gas should be re-accreted into \nqh\ after the collision.  
 
  Numerical simulations of galaxy tidal interactions or mergers (e.g. Hibbard \& Mihos 1995,    
Barnes \& Hernquist 1996,  Mihos 2001) have shown that most of the material in stellar 
and gaseous tails remain bound to a parent disk, hence falling back towards it.
There are evidence in the velocity field that  gas in the filaments is dynamically
bound to \nqh. Indeed, this study confirms that no extremely high velocity departures with  
respect to the systemic velocity are observed in the filaments
(see also Combes et al. 1988, \ky). The highest velocity  amplitudes (in absolute values) are  found in the disk.
Moreover, the filaments  seem to participate to the rotation of the galaxy, with redshifted (blue-shifted) gas 
to the North (South respectively) and with velocities close to the systemic value 
in parts of F1 and F2 that roughly lie along the minor axis.  
Though F0 is obviously less redshifted than part of F1 while it is the northernmost filament, 
this trend  exactly follows the disk rotational pattern.   
The diffuse filaments of \nqh\ are thus apparently bound to it and it is likely that 
they will be reaccreted towards the disk. 
 
 \begin{figure*}
\includegraphics[width=17.2cm]{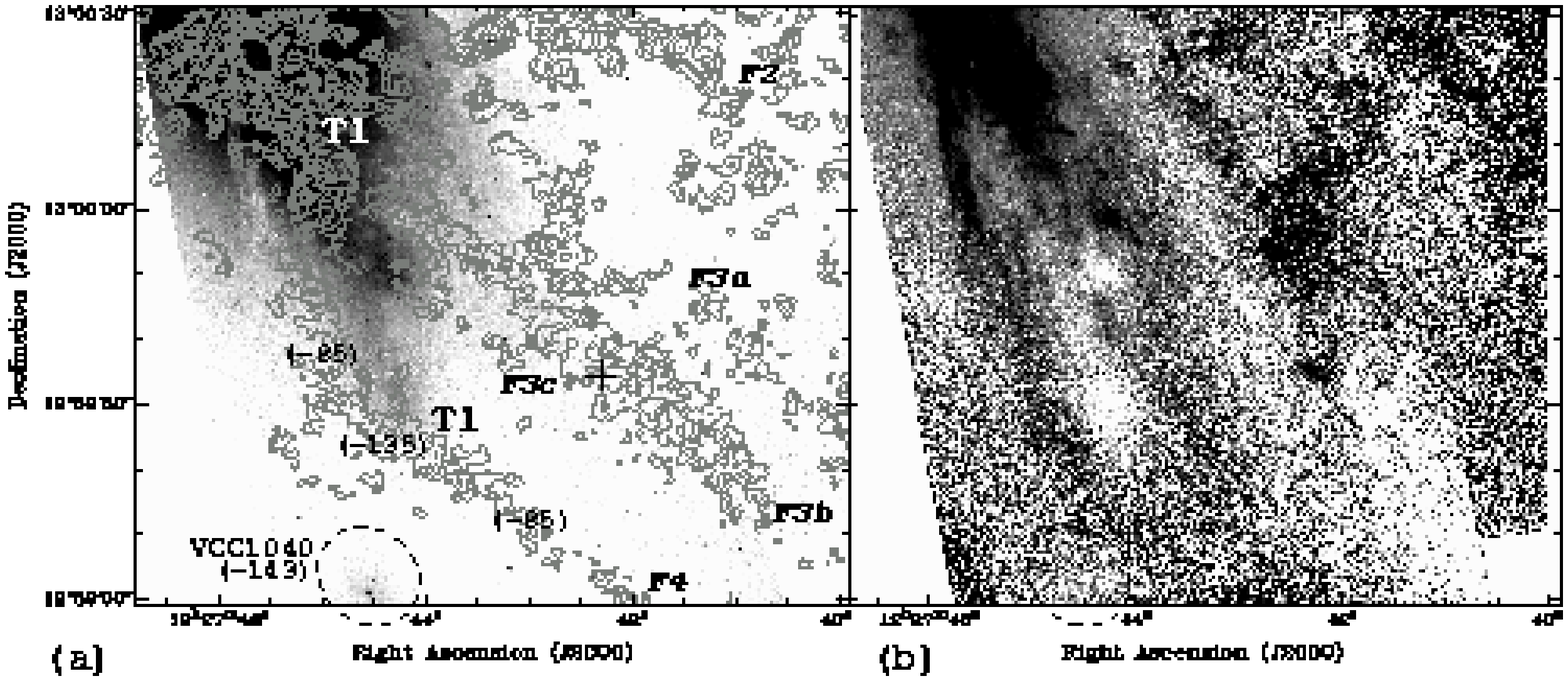}
\caption{ \textbf{(a)}. F450W WFPC2-\emph{HST} image showing \vcc\ and the southern parts of
\nqh. Full-line contours represent the \ha\ emission in \nqh, a dashed line contour an isophote of \vcc. Labels F2 to F4 refer to
the \ha\ filaments. The opposite extremities of the stellar tail T1 are delineated
by labels T1. Numbers into brackets represent radial velocities in F4 and of \vcc\ (in \kmps). 
 A cross  marks the location of the object 2MASXiJ1227422+125935 (see text).
\textbf{(b)}. F450W-F814W colour index map. Dark shades represent dust lanes. The dark area to the East of \vcc\ 
is \emph{not} due to dust but to the sky background level.}
\label{finter}
\end{figure*}

\subsection{Ram pressure stripping}
\label{stripicm}

Signatures of ram pressure stripping (hereafter RPS) have already likely been observed in nearby groups and clusters galaxies 
in the form of linear structures, filaments, off-plane ring-like and comet-like morphologies of either ionized or neutral gas  or radio
continuum (Gavazzi et al. 2001, Vollmer et al. 2000, Bureau \& Carignan 2002, Vollmer et al. 2004a).
In the Virgo cluster, there exist observational indications of ongoing or recent RPS.
The most probable cases are among the most \hi-deficient galaxies, like 
  NGC 4388 (Veilleux et al. 1999, Yoshida et al. 2002, Vollmer  \& Huchtmeier 2003) 
  or NGC 4522 (Vollmer et al. 2000, Kenney et al. 2004, Vollmer et al. 2004a).
A case where a RPS could happen with a gravitational interaction has also been presented 
for NGC 4654 (Vollmer 2003).

It has been claimed that a tidal interaction and an ISM-ISM collision with \nqc\ were 
probably responsible for the displacement of  \hi\ and part of the CO gas outside 
the galaxy plane of \nqh\ and for the formation of the ionized gas filaments  (Combes et al. 1988, \ky). 
\nqh\ is yet a good candidate for an active RPS by the Virgo ICM (Kotanyi et al. 1983, Cayatte et al. 1990, 1994, \ky, Vollmer et al. 2004b).
The fact that  F2, F3, parts of F0 and F1 and the second half of F4 are all directed along the NE-SW direction
could be an indication that a common stripping process is responsible for their shaping and  high degree of collimation.  

 A key-point for a RPS hypothesis is to explain why no gas is observed
  at high  velocities relative to the galaxy in the \ha\ filaments. This observational fact is not expected because
  of the rapid l-o-s relative motion of \nqh\ with respect to the cluster ($\sim 1000$ \kmps).
  If high velocity gas really exists, which reasons could explain its non-detection ?  
  Firstly, we recall that our restricted spectral ranges  prevent from detecting any 
  extremely high velocity departures w.r.t. the systemic velocity. 
 If present, detecting it would need other observations performed at other wavelengths or with a larger spectral band. 
However, 
 the Fabry-Perot spatial distribution of \ha\   matches well the one obtained from deep \ha\ $+$ \nii\ images, showing 
 that no gas is missed by our observation. In other words, it seems implausible that a large amount of gas 
  at high l-o-s velocity exists in the optical  filaments.
  Then, numerical simulations of RPS could help to answer this question.
 Major improvements on the modelling of RPS have been done recently (Abadi et al. 1999, Quilis et al. 2000, Schulz \& Struck 2001, Vollmer et al. 2001) 
 and models succeeded in reproducing kinematical properties and  morphological anomalies of cluster galaxies (e.g. Vollmer et al. 2004b).
For instance, the case of NGC 4388 is particularly interesting because it is another highly inclined galaxy in the cluster core.
A RPS model for NGC 4388 (Vollmer  \& Huchtmeier 2003) has shown that most of the gas remains bound 
to the galaxy and that gas particles with the highest velocities w.r.t. the galaxy appear at high distance from the disk (few tens of kpc). 
The important  parameters of such simulations are the ICM density, the galaxy velocity  relative to the ICM (2000 \kmps\ for NGC 4388) and  
the inclination angle of the galaxy disk to its orbital plane. 
Although \nqh\ is orbiting in the Virgo cluster under different conditions than NGC 4388 and  could have been considerably disturbed by a massive companion,
one could expect that most of its gas resides close the galaxy systemic velocity within the first tens of kpc from the centre, as  is observed.
If a small fraction of gas with high relative velocity exists, it could be now located further away from the nucleus,
 outside the  field of view. 
 A good test would be to search neutral gas in a very large field-of-view around the galaxy. 

 Perhaps another good test for an ongoing RPS scenario would be to study the stellar kinematics of the southern region, 
 where filaments F3 and F4 appear to superpose on large and small (respectively) stellar tails.  
 Since RPS only acts on ionized and neutral gas and not on the stars, it is expected that a 
 kinematical decoupling  has developed  between gas and stars.
 It should however be remembered here that simulations of galaxy collisions have already shown velocity differences reaching $\sim 50$ \kmps\
 between stars and gas in tidal tails (Mihos 2001), though their global evolution remains coupled (re-accretion on the parent galaxy disk, ...). 
 We do not know the amplitude of an hypothetical velocity difference between stars and gas in the concerned regions of NGC 4438 
 that would be caused by a tidal interaction \emph{alone}. Numerical models of RPS coupled with a tidal interaction 
 should help to disentangle their respective role on the evolution of the filaments. 

\subsection{A perturbed region to the South}
\label{inter2}
\subsubsection{The peculiar Filament F4}
Filament F4  is probably the most intriguing observed filament. 
A $B$-band image taken from the GoldMine database and F450W, F814W
WFPC2-\emph{Hubble Space Telescope} images taken from the \emph{HST} archive are used
to illustrate the tight relation
that exists between stars, gas and dust around   Filament F4
and the outer southern part of the disk (Figs.~\ref{finter} to~\ref{clump}). 
The  WFPC2-\emph{HST} images were reduced using the 
STSDAS package in \textsc{iraf}. Warm pixels and cosmic ray hits were removed using the  \emph{warmpix} and \emph{crrej} tasks 
and the final mosaics were created using the  \emph{wmosaic} task. Cosmic rays were also treated using the \emph{lacos\_im} procedure (van Dokkum 2001).
We also built a F450W-F814W ($\simeq B-I$) colour index map from the ratio of  F450W to F814W (Fig.~\ref{finter}b) in order
 to follow the dust lanes and star formation regions across and outside the galactic disk. 
Basically, redder colours appear here as darker shades and correspond to dusty regions (dust lane associated with the disk and
filaments, and the western dust layer) while bluer colours appear as lighter shades and can correspond to star formation sites.
The colour index image principally shows that the dust is along a prominent ring-like structure in the inner parts of the galaxy and
also present outside the disk to the West and South. \hii\ regions along the filaments have colours as blue as 
$m_{\rm{F450W}}-m_{\rm{F814W}} = -1.3$ mag.

We notice here the presence of an irregular-shaped object with a colour of up to 1.2 mag, lying along the edges of the prominent dust lane  and  Filament F3b
(R.A. $\sim$ 12$^h$ 27$^m$ \fas{42}{3}, Dec. $\sim$ +12$\degr$ 59\arcmin\ \as{35}{2}). It is nearly aligned with the bright warped stellar structure of the southern
edge of the disk and Filament F3c (Fig.~\ref{finter}a).
The only referred object closest to this location is called 2MASXiJ1227422+125935 and is classified as a galaxy of diameter \am{1}{2}
in the 2MASS database (Jarrett et al. 2000). We deduce an upper limit ``diameter" 
of $\sim 6$\arcsec\ from the F814W-\emph{HST} image, which is considerably lower than
the referred value.  No gas is detected within this object and because its redshift is unknown, it is not yet possible to determine whether
it is related to \nqh\ or whether it is a background galaxy. 
 
As seen in Figs.~\ref{finter}a and b, a first half of F4 lies along the eastern edge of a stellar tail (hereafter T1)
and exactly superposes on a dust lane. Then F4  bends  towards the SE from the tip of T1.  
The northernmost part of T1 seems to arise from the inner regions of the disk. 
A lower limit for the length of T1 is 1\arcmin\ (4.7 kpc). 
It is shorter than the two large-scale outer tails, which extend on distances  of at least \am{4}{1} (or 19 kpc) towards the NE
 and \am{3}{1} (or 14.5 kpc) towards the SW.
T1 is obviously crossed by a brighter and warped stellar tail (R.A.$\sim$ 12$^h$ 27$^m$ 44.5$^s$,
Dec.$\sim$ +12$\degr$ 59\arcmin\ 55\arcsec, Fig.~\ref{finter}a). 
Figure~\ref{finter}a  shows that this bright warped stellar tail arises from the SE near
side of the galaxy disk. T1 then lies straight towards the South, pointing in the direction of \vcc, a  dwarf elliptical galaxy.
It has a lower surface brightness than the stellar disk and the bulge/halo of \nqh.
The  tips of T1 and the brighter warped tail have blue colours ($-0.2$ mag. on average).
 T1 is reminiscent of tidal stellar streams like those 
observed around massive spirals such as the Milky Way (Ibata et al. 2001a) or Messier 31 (Ibata et al. 2001b). 
   The next section presents arguments to explain its formation by the presence of \vcc. 

\subsubsection{A role for \vcc\ ?}
\label{inter3}

 \begin{figure}
 \centering
  \includegraphics[width=7cm]{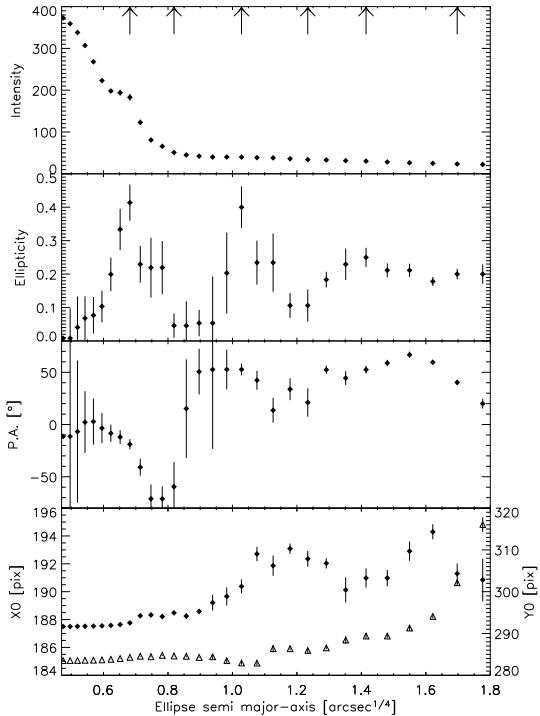}
  \caption{Isophotal analysis of \vcc\ from a F814W WFPC2-\emph{HST} image. From Top to Bottom: intensity, ellipticity, position angle (in degree),
  central X and Y coordinates (in pixel, full dots and open triangles respectively) variations as function of ellipse semi-major axis (in arcsec$^{0.25}$). The arrows indicate the radii 
  of the isophotes displayed in Fig.~\ref{clump}.}
  \label{ellivcc}
\end{figure}

\vcc\ is a dwarf-nucleated elliptical galaxy and is the closest companion of
\nqh. It is projected at only 
\am{1}{5} (7 kpc) from the nucleus of \nqh\ (to be compared with \am{4}{5} or 21 kpc between 
\nqc\ and \nqh) and $\sim 25\arcsec$ (2 kpc) from F4. 
It has a redshift of $-143 \pm 63$ \kmps\ (Conselice et al. 2001) 
which coincides within the errors with the velocity of the region where F4 bends 
and with the mean radial velocity of F4. 
 Though dwarf elliptical galaxies are the most common type of galaxies 
found in the local galaxy clusters like Virgo  (Ferguson \& Binggeli 1994) and the probability of
finding foreground or background dEs in the direction of the cluster  core is high, 
their proximity in sky position  and redshift  likely makes \vcc\ a genuine companion of \nqh.  
 Its colour index $B-R$ is  1.6 (\ky) and indicates an old object (i.e. with an age $>> 100$ Myrs),  
which is usual for dEs (Rakos et al. 2001, Conselice et al. 2003, Jerjen et al. 2004), and no ionized gas is found in it. 

 \ky\ have noticed the presence of several apparent stellar clump debris including \vcc, 
which could have been formed during the interaction between \nqc\ and \nqh.  
 Tidal dwarf galaxy candidates are usually blue, gas rich, young and irregular-typed 
structures always observed along stellar tidal tails (e.g. Duc et al. 1997), which properties do not match with those of \vcc.
Morever the time-scale invoked to form a dE is several hundred Myrs (Kroupa 1998), which  is shorter than the elapsed time 
since the interaction between \nqc\ and \nqh\ ($\sim$ 100 Myrs according to the Combes et al.' simulations). 
It seems unlikely for \vcc\ to be a tidal dwarf galaxy.  

 Several elements show that the environment around \vcc\ and \vcc\ itself are perturbed. 
 Firstly, an \ha\ emission is found between Filament F4 and the dwarf galaxy (Fig.~\ref{finter}a).
 This emission actually appears to be an unusual ionized gas loop to the East of the F4 bending region.
 
\begin{figure}
  \resizebox{\hsize}{!}{\includegraphics{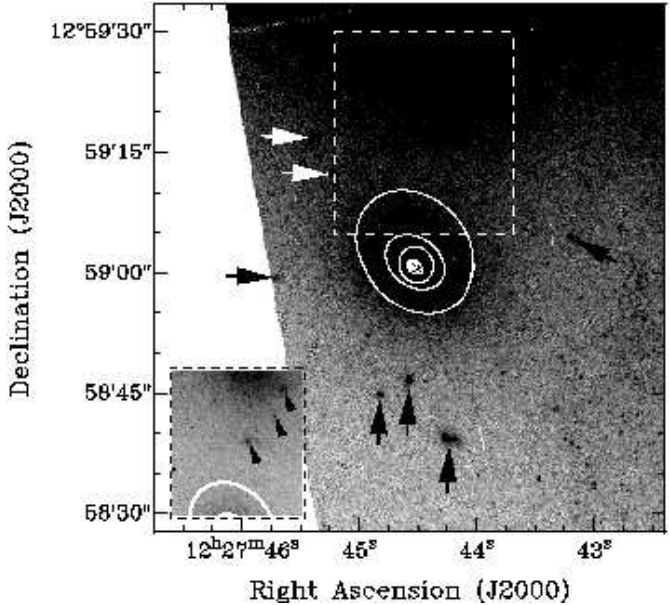}} 
  \caption{Enlargement of an F814W WFPC2-\emph{HST} image around \vcc. The arrows indicate different clumps surrounding the dwarf galaxy while ellipses 
  indicate fitted isophotes of \vcc\ to the image at the radii flagged by arrows in Fig.~\ref{ellivcc}. The bottom-left insert is a focus of the white dashed box
  for another intensity scale to emphasize on brighter clumps \textit{between} the dE and the tip of the stellar tail T1.}
  \label{clump}
\end{figure}

 Secondly, the isophotal analysis of the dE light distribution reveals its perturbed morphology, as shown in Figs.~\ref{ellivcc} and~\ref{clump}.
 Ellipses of variable geometrical parameters (centre coordinates X0 and Y0, ellipticity 
 and position angle of the isophotes) were fitted to a F814W WFPC2-\emph{HST} image. 
  A lower limit of the \vcc\ radius is thus $\sim$ 10\arcsec\ (780 pc), i.e. for the last
 accurately measured isophote. \ky\ found an outer radius of 14\arcsec.
 At higher radius than $10\arcsec$, the light distribution is partly contaminated by
 the diffuse stellar light of \nqh\ and by dust to the N and NW outer parts of the dE, 
 hence results are not shown for these radii due to too large uncertainties on the parameters. 
 The graphs show that there is a substantial offset of the position of the photometrical centre as the radius increases
 and that the Y-offset of the outer ellipses is larger than the X-offset.
 As an example,  the isophote with semi-major axis $r \sim$ \as{8}{4} (last arrow) has its centre offset by
 $\sim$ \as{1}{8} (140 pc) to the N-NW direction with respect to the innermost isophotes, within a
 ratio of $\rm\frac{Y-offset}{X-offset} = 6$.
 Moreover, the galaxy  is less rounder in its outer parts ($\epsilon \rightarrow$ 0.2) than in its inner ones
 ($\epsilon$ $\sim$ 0.05 for \as{0}{4} $\leq r \leq$ 1\arcsec),
 though passing through a significant peak at r=\as{1}{1} ($\epsilon$ = 0.4). 
 Twists of the isophotes are also clearly detected.
 The $P.A.$ indeed drastically varies  between $\sim 10\degr$ and $70\degr$ (for $r > 0.5\arcsec$).
  At a radius of $\sim$ \as{0}{25}, the intensity bump, the ellipticity peak and the drastic 
 change of photometric $P.A.$ are likely due to the nucleus of the dwarf. 
 Figure~\ref{clump} displays some isophotal ellipses (which semi-major axis positions
 are flagged by arrows in Fig.~\ref{ellivcc}) and illustrates well  the measured asymmetry of the dE. 
 
 Finally,  numerous clumps are observed in the close vicinity of \vcc\ as well as between the dwarf galaxy and the tip of T1 (Fig.~\ref{clump}). 
  Their colour range is large, from blue to red clumps ($-1$ to $2$ mag) indicating that they could be  gas rich objects to old globular clusters.
  No gas was detected in them and their radial velocities are  unknown   so that their possible 
  association with the southern region remains to be confirmed.
  Other knotty structures are found within the NE and SW large-scale stellar tidal arms (\ky).
  It is surprising to find stellar clumps all around the dE,
   because they are not directly associated to the large-scale tails. Potential reservoir of stars for them could thus be \vcc\ or T1.

 Which mechanism could explain all these perturbations~?
We  emphasize here  on a minor tidal interaction event that could have occurred 
(or perhaps that would  be occuring) between \vcc\ and \nqh, due to the presence of the dust lane and of the stellar tail (T1) 
 both pointing towards \vcc\  and to the fact that the \vcc\ redshift is comparable with velocities 
 observed along the gaseous filament F4, and thus perhaps with the velocity of the tail.  
A comparison of the \vcc\ tidal radius $r_t=R\left( \frac{m}{3M} \right)^{\frac{1}{3}} \sim 7 \arcsec$ (0.5 kpc) 
 with the \vcc\ radius of  $r \ga$ 10\arcsec\ (0.78 kpc) 
implies that a tidal stripping of its external parts by the gravitational forces of \nqh\ could be ongoing. 
We have used here   $R \sim 7$ kpc as the distance between the two galaxies, $m \propto 10^8$ $\msol$ as  the mass of \vcc, which is typical of dwarf ellipticals, 
$M \propto 10^{11}$ $\msol$ for \nqh\ (Combes et al. 1988)\footnote{The estimated value of the tidal radius is a lower limit because of the use of 
the \emph{projected} distance of the dwarf to the nucleus of \nqh.}.    
 Early  simulations of the encounter with \nqc\ (Combes et al. 1988) did not predict T1
  but only the large-scale stellar asymmetry of \nqh\ and one has to imagine another triggering event for its formation. 
The stream T1 and the stellar concentrations around the dwarf galaxy 
 can be consequences of such a minor tidal event. 
Photometrical anomalies are often observed in
 dwarf early-type galaxies of the Virgo cluster (Binggeli et al. 2000, Barazza et al. 2003) 
 and it is still unclear whether they are due to intrinsic or external perturbations to the dwarf galaxies. 
 We think that the asymmetries measured in \vcc\ are of the same nature as those measured in other dEs. 
 An interaction with \nqh\ could also explain the photometrical disturbance of \vcc,  by stretching the outer parts towards \nqh,  
 as measured from the isophotal analysis. 

 This would not be the first time that a minor interaction opposing two galaxies, one large and one dwarf, 
is observed in the Virgo cluster. Other possible cases  have indeed been presented with 
the couples M49-VCC 1249 (Sancisi et al. 1987, McNamara et al. 1994), NGC 4694-VCC 2062
 (van Driel \& van Woerden 1989, Hoffman et al. 1993) and  M86-VCC 882 (Elmegreen et al. 2000).  
A detailed   analysis of the stellar populations in the tails and in the galaxies would be very useful to test the minor interaction hypothesis. 

\section{Conclusions}

New optical observations in the \ha\ emission-line of the Virgo cluster galaxy \nqh\ were presented, as part of a large survey dedicated to the kinematics
of Virgo spiral galaxies. Using Fabry-Perot interferometry data at an effective angular resolution of $\sim 2\arcsec$, the optical velocity field of the galaxy was mapped for the first time.  
\nqh\ is one of the most disturbed Virgo spirals which has probably undergone a tidal interaction with the massive companion \nqc, and which is 
probably undergoing the effects of ram pressure stripping by the hot ICM in the cluster core.
The \ha\ velocity field reveals the complex gas kinematics within the galactic disk, the nuclear shells and the external filaments.  
Our conclusions can be summarized as follows:

(1) The nuclear shell of \nqh\ is likely expanding, as revealed by the multiple spectral components detected in the FP data-cube along the NW shell. 
Further observations of the nuclear region at higher spectral resolution and larger spectral coverage than our observations are needed
 to investigate the internal motions of the nuclear outflow.

(2) A new filament is observed to the North of the disk. It is short, clumpy and has no X-ray counterparts  
while the other filaments are much more extended, diffuse and exhibit a X-ray counterpart. 
 Two velocity components are observed in some pixels of the data-cube in the approaching side of the disk. 
 One of the component has a redshifted kinematics which totally contrasts with the local, blue-shifted velocity field of the approaching half. 
Its velocity is however consistent with that found in nearby, diffuse, off-plane knots, suggesting that both could belong to a same morphological structure.    
 The global kinematics of the extended filaments indicate that they are likely bound to the galaxy. 

(3) A tight relation between the presence of a small stellar tail aligned with a dust lane and an \ha\ filament South of the galaxy,
and of the dwarf elliptical galaxy \vcc\ is outlined. The gaseous filament suddenly changes its orientation at the tip of
the  stellar stream. The \vcc\ redshift is comparable with velocities observed in the \ha\ filament.
We suggest that a minor tidal interaction has occurred, or is perhaps  occurring, between \nqh\ and the dwarf galaxy, 
causing the disruption of the small galaxy and the formation of the stellar tail.

Further 3D numerical simulations taking into account both  tidal interactions and ram pressure stripping would be crucial   
to improve our understanding of the evolution of this complex  system.

\acknowledgements

We thank Marie Machacek for gratefully providing with the \emph{Chandra} X-ray image of the galaxy and for interesting discussions, 
Anthony Moffat for a careful reading of the manuscript, Olivier Daigle for his technical support and the anonymous referee for constructive remarks.
This work was partly funded by the grant Regroupement Strat\'egique -
Observatoire du mont M\'egantic of FQRNT (Qu\'ebec) and by the Minist\`ere de 
l'\'Education Nationale, de la Recherche et de la Technologie (France). We made use of GoldMine - 
Galaxy On Line Database Milano Network (http://goldmine.mib.infn.it)
and the \textit{Hubble Space Telescope} data archive (http://archive.stsci.edu/hst).

\end{document}